\documentclass[apjl]{emulateapj}
\usepackage{hyperref}
\usepackage{color}
\hypersetup{colorlinks,
 citecolor=blue,
 linkcolor=blue}
\usepackage{natbib}

\usepackage{hyperref}
\hypersetup{colorlinks,%
 citecolor=blue,%
 linkcolor=blue}
\definecolor{lightblue}{rgb}{.70,.95,1} 
\def\aj{AJ}
\def\apj{ApJ}
\def\apjl{ApJ}
\def\apjs{ApJS}
\def\aap{A\&A}
\def\mnras{MNRAS}
\def\nat{Nature}
\def\araa{ARA\&A}

\shorttitle{An isolated cE with AGN}
\shortauthors{Paudel et al}
\begin{document}

\title{SDSSJ085431.18+173730.5: The First Compact Elliptical Galaxy Hosting an Active Nucleus}

\author{Sanjaya Paudel\altaffilmark{1}, Michael Hilker\altaffilmark{2},
 Change Hee Ree\altaffilmark{1},  Minjin Kim\altaffilmark{1}
  }
\affil{
$^1$ Korea Astronomy and Space Science Institute, Daejeon 305-348, Republic of Korea\\
$^2$ European Southern Observatory, Karl-Schwarzschild-Str. 2 D-85748, Garching bei Munchen, Germany\\
}

\altaffiltext{2}{Email: sjy@kasi.re.kr}

\begin{abstract}
We report the discovery of a rare compact early-type galaxy, SDSS J085431.18+173730.5 (hereafter cE\_AGN). It has an half light radius of R$_{e}$ =  490 pc and a brightness of M$_{r}$ = $-$18.08 mag. Optical spectroscopy available from the Sloan Digital Sky Survey (SDSS) reveals the presence of prominent broad-line emissions with the H$\alpha$ broad component width of FWHM=2400 km/s.  The black hole (BH) mass, as estimated from the luminosity and width of the broad H$\alpha$  emission, is 2.1$\times$10$^{6}$ M$_{\sun}$. With the help of surface photometry, we perform a detailed analysis of the structural properties. The observed light distribution is best modeled with a double S\'ersic function. Fixing  the outer component as an exponential disk, we find that the inner component has a S\'ersic index of n = 1.4. Considering the inner component as bulge/spheroidal we find that cE\_AGN remains consistent in both  BH mass -- bulge mass relation
and BH mass -- bulge S\'ersic index relation. Given these observational properties, we discuss its possible origin investigating the surrounding environment where it is located.

\end{abstract}


\section{Introduction}
In the nearby Universe, compact dwarf galaxies are a rare class of objects. The majority of low-mass early-type galaxies (preferentially called dwarf ellipticals, dEs) do not follow the near-linear Log(size)-magnitude distribution defined by giant ellipticals but have nearly constant sizes over large luminosity ranges  \citep{Janz08,Kormendy09}. A continuous relation joining bright and faint galaxies exists, but this is more complex than a simple power law  \citep{Graham03,Graham13,Ferrarese06,Janz08}. There exist a very compact, high-surface brightness and non-star-forming elliptical galaxy \citep[cE,][]{Chilingarian09,Faber73}, which seems fall on the extension of the relation defined by massive ellipticals  (Es) alone \citep{Kormendy09,Misgeld11,Paudel14}.

M32 and VCC1297 are classical examples of a cE in the nearby Universe \citep{Faber73}. These compact objects were thought to be extremely rare. For example, in the Virgo cluster there are only a handful of possible cE candidates while more then 500 dEs are confirmed members by redshift measurements. An increasing number of cE populations located in different environments have been reported in recent works  \citep{Chilingarian09,Chilingarian15,Huxor11,Huxor13,Mieske05,Paudel14,Price09}. Most of them, however, are found in dense environments, particularly around massive galaxies. Such proximity to a massive host favors a scenario in which they arise through the tidal stripping of larger galaxies, where the inner bulge component becomes a naked compact galaxy after losing its outer disk component during the interaction with its massive host \citep{Bekki01,Choi02}.  Indeed, the discovery of cEs with tidal debris near a massive host confirms this picture \citep{Huxor11}.

\begin{figure}
   \includegraphics[width=8cm]{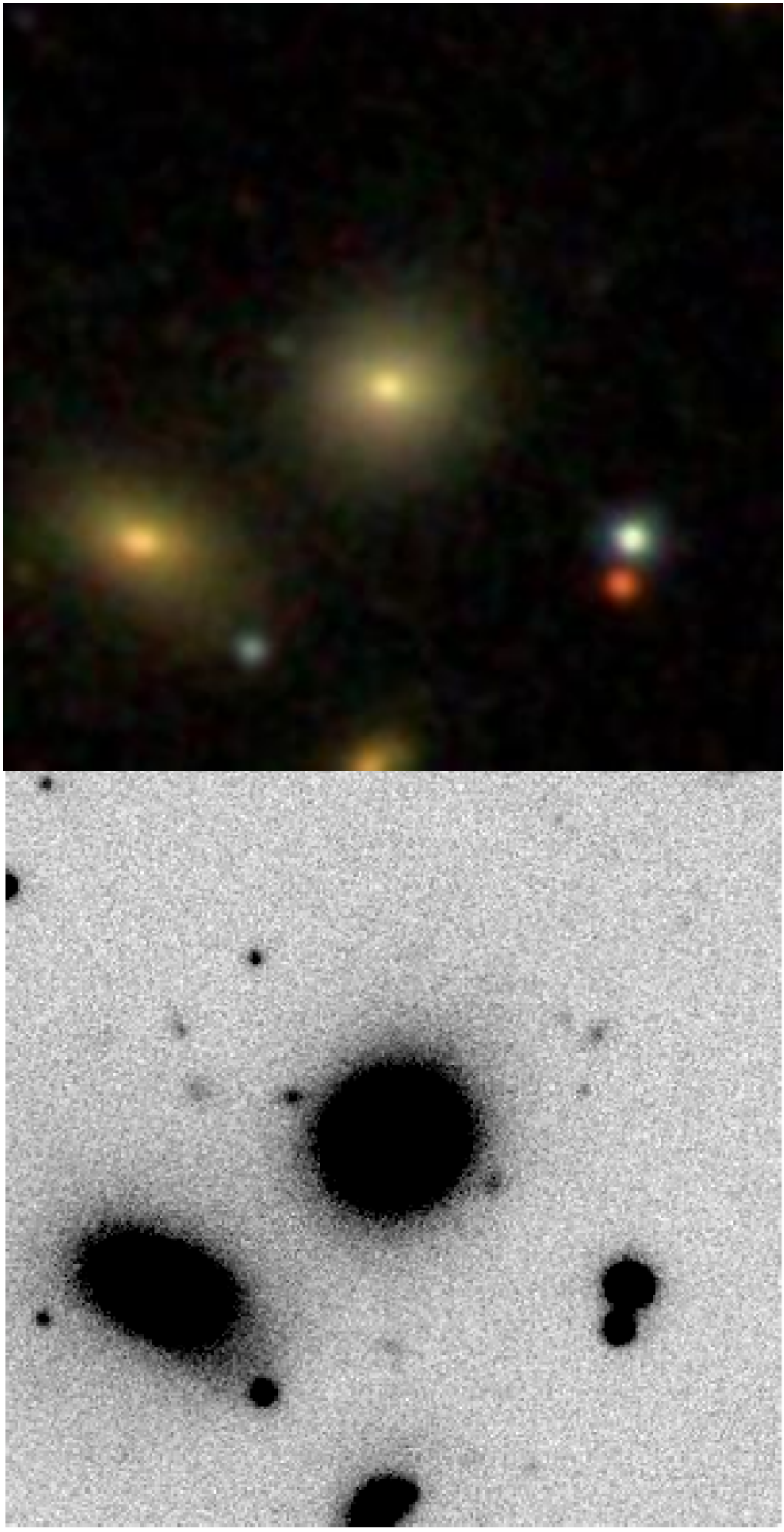}
  \caption{Optical outlook of cE\_AGN within a field of view 50"$\times$50" centered at RA = 133.62990 and Dec = 17.62513. The SDSS g-r-i color composite 
  and CFHT $i-$band images are shown in the top and bottom panel, respectively. }
  \label{gpic}
   \end{figure}

 \begin{table*}
 \caption{ Global properties of SDSS J085431.18+173730.5 }
\begin{tabular}{lccccccccccc}
\hline
Galaxy       &  M$_{k}$ &  M$_{r}$ & R$_{e}$&n & $<$$\mu_{r}$$>$& $g-r$  & z  & Age & log(Z/Z${\sun}$)  & M$_{\bullet}$ & M$_{*}$\\
    &     mag &  mag  & pc & & mag/arcsec$^{2}$ & mag & & Gyr & dex & log(M$_{\sun}$)& log(M$_{\sun}$)\\
\hline
cE\_AGN  &  $-$20.03 & $-$18.08& 490  &1.4 & 19.36 & 0.83 &  0.0140  & 9.95$\pm$0.4 &$-$0.12$\pm$0.01 & 6.3   & 9.2 \\   
CGCG 036-042 (P14) & $-$20.78 & $-$18.21 &559 & 1.7 &19.64 & 0.77 & 0.0068 & 7.15$\pm$1.2 & $-$0.18$\pm$0.07 & & 9.5\\
\hline
\end{tabular}
\\
\\
 The global photometric parameters - total brightness and half light radius-- are derived using Petrosian photometry where the galactic extinction is corrected using \cite{Schlafly11}. We analyzed the SDSS-optical spectroscopy to derive stellar population parameters and BH mass, see the text. 
\label{htb}
\end{table*}

Galaxies follow several scaling relations between their total mass and structural, kinematic and stellar population properties. It has been argued, with observational support, that supermassive black holes (SMBHs) are present at the center of galaxies, and a correlation between the masses of SMBHs and host galaxies bulge masses exists. However,  there has been a debate on the universality of this relation at the low mass end. It is also not conclusive whether a majority of dwarf galaxies possess BHs at their centers, in addition to that most dwarf galaxies are bulge-less  (or pseudo-bulge) galaxies. Instead, nuclear star clusters (NSCs) are common in low mass galaxies galaxies \citep{Cote06,Georgiev14,Graham03}. \cite{Ferrarese06} show that the scaling relation between BH mass and bulge mass can be extended to the low-mass regime when taking the NSC mass and total host galaxy mass instead, but see also \cite{Graham14}.

\cite{Paudel14}, hereafter P14, report a cE galaxy located in an isolated environment that possesses an extended low-surface brightness feature; a likely sign of tidal interaction. In this work, we report the discovery of another cE type galaxy, SDSS J085431.18+173730.5, located in a nearly-isolated environment that possesses a prominent broad-line emission in the optical spectrum; a likely sign of nuclear accretion by the central BH. In this work we assume a cosmology with H$_{0}$ = 71 km s$^{-1}$ Mpc$^{-1}$, $ \Omega_{M}$ = 0.3, and    $\Omega_{\Lambda}$  = 0.7.

\section{The data analysis}
SDSS J085431.18+173730.5 (hereafter  cE\_AGN) has been serendipitously discovered in SDSS-DR12 \citep{Alam15} in the course of  searching for low-luminosity early-type galaxies in isolated or low density environments. It is located at z = 0.014 with no significantly massive neighbor. Based on NED query, the nearest companion is a star-forming dwarf galaxy (M$_{r}$ = -17.33 mag) which is located at a sky-projected distance of 279 kpc in south-east direction and the relative line-of-sight velocity between cE\_AGN and its companion is 194 km/s whereas ARK 186 is the nearest giant (M$_{r}$ = -19.82 mag) neighbor galaxy. The sky-projected physical distance between ARK 186 and cE\_AGN is 639 kpc and the relative line-of-sight velocity between them is 38 km/s. This part of the sky, within the redshift range 0.011 to 0.019, is sparsely populated by galaxies where we find only 13 galaxies within the radius of one degree (corresponding to a physical radius of 1.01 Mpc) centered on ARK 186. The NGC 2672 group, with  z =  0.014487,  at a sky-projected distance of $\sim$2Mpc from cE\_AGN is the nearest large concentration of galaxies.

\begin{figure}
   \includegraphics[width=8.5cm]{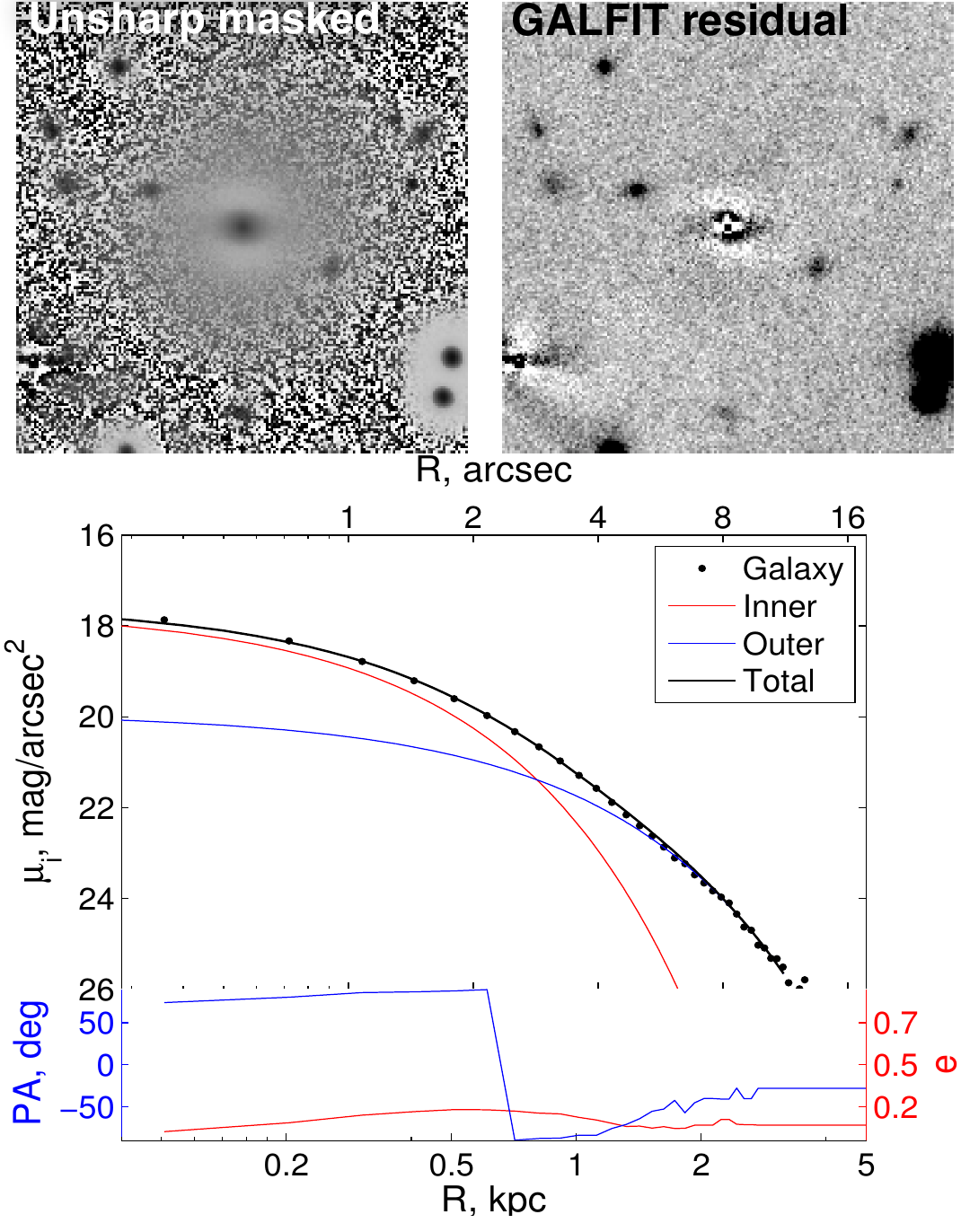}
  \caption{The unsharp-masked and GALFIT residual images are shown in the left and right top panel, respectively. IRAF $ellipse$ output parameters -- azimuthally averaged surface brightness, position angle and ellipticity -- are shown in the bottom figure where the observed major axis light profile (black dots) is fitted with a double S\'ersic function. The solid black line represents the best fit combined S\'ersic function. }
  \label{galfit}
   \end{figure}

\begin{figure*}
   \includegraphics[width=16cm]{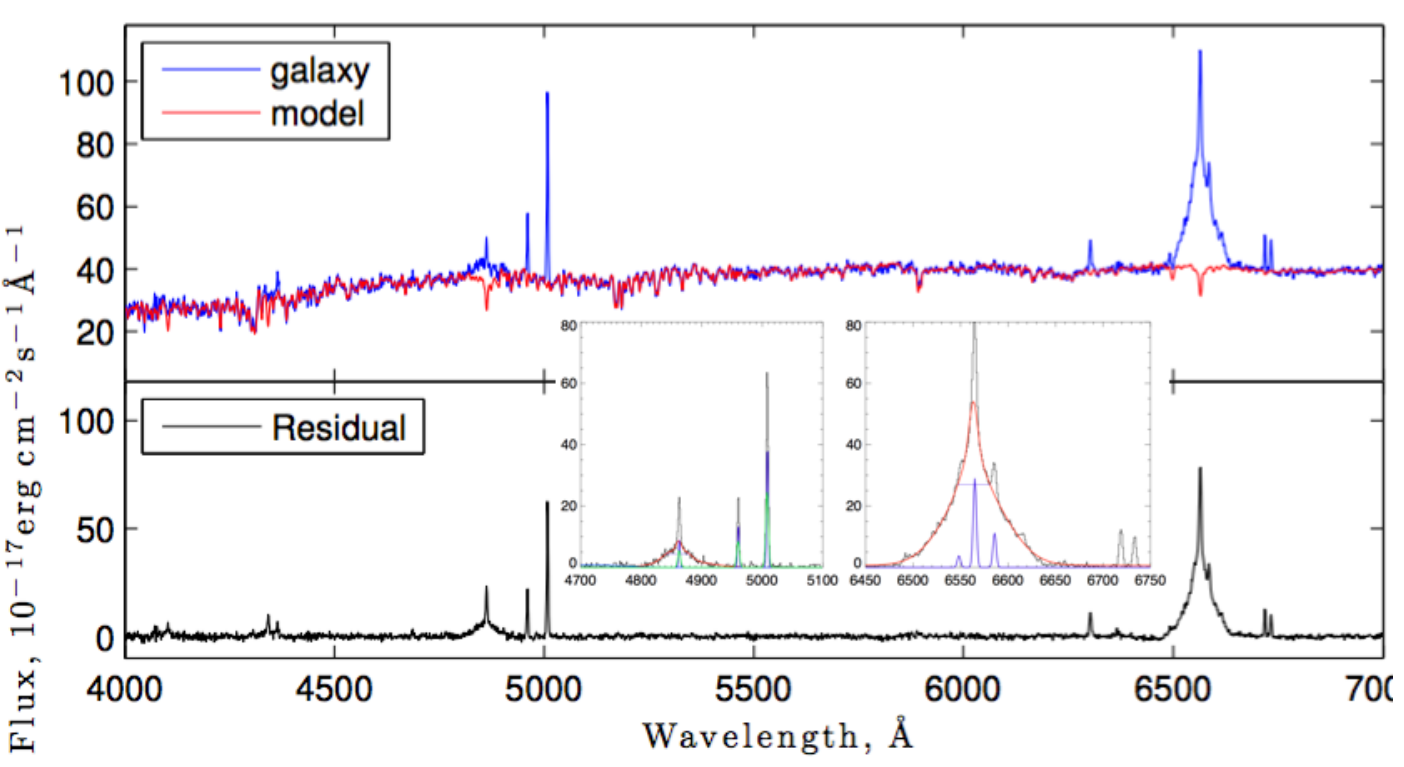}
  \caption{ The SDSS optical spectrum of cE\_AGN (blue), together with its best-fitting SSP model spectrum (red). The fit residuals are shown in the lower panel. The fit is generally consistent within 5 per cent of the observed flux. In the inset, we show modeling of the broad emission lines with multi component gaussian profiles, see text for more detail.}
  \label{sspfits}
   \end{figure*}

 \subsection{Imaging}
We primarily made use of the SDSS-III \citep{Ahn12} imaging data and catalog to explore the parameter space in searching for compact objects that we are interested in. We present the an optical images of cE\_AGN in Figure \ref{gpic}, a $g-r-i$ color composite taken from the SDSS skyserver at the top and a CFHT $i-$band image at the bottom. We find that the SDSS image quality is rather poor with an average seeing in $r-$band  of $\sim$1{\arcsec}. The SDSS measured half-light radius of cE\_AGN is 1.98{\arcsec}. Therefore, to avoid any uncertainty in the derived structural parameters due to the resolution limit, we searched for higher quality images in other publicly available archival imaging databases. We found higher quality imaging data in the Canada France Hawaii Telescope (CFHT\footnote{http://www.cfht.hawaii.edu}) archive, with an average seeing of 0.6", which was observed in the $i-$band.

 We acquired the pre-processed, level 2 calibration, fits files which include dark current subtraction, flat normalization and the magnitude zero-point supplemented in the header. The observations were made as a part of the CFHT Very Wide survey. With the seeing of 0.6{\arcsec}  and a pixel scale of 0.18{\arcsec}/pix, CFHT data provides significantly higher image quality compared to the SDSS observations. However, note that the CFHT MegaCam images are known to have severe issues with sky background subtraction due to the mosaic pattern of multiple CCD chips. Thanks to its compactness, our object extends entirely in one chip,  thus sky-background subtraction becomes straightforward.

As we can see in Figure \ref{gpic}, a background (z = 0.105) galaxy is located next to cE\_AGN. We removed the background contamination by subtracting its light from the fits image using the IRAF\footnote{http://iraf.noao.edu} $ellipse$ and $bmodel$ tasks. An unsharp masking technique has been applied to examine the presence of any substructural features such as spiral arm/bar or inner disk in cE\_AGN.  Finally, we used  {\sc galfit} \citep{Peng02} to carry out a 2D multicomponent structural analysis where we modeled the light distribution with several different possible combinations, i.e one with a simple S\'ersic function and the other with a double S\'ersic function with or without a central gaussian point source. For the double S\'ersic function, we fixed the outer component to be an exponential, i.e., $n = 1$.  A PSF image, required by  {\sc galfit}, was prepared by stacking the stars found in the same field of view. Foreground and background sources were masked by providing a mask image that we prepared manually.

With visual inspection of these residual images, we find that the use of a double S\'ersic function provides a better fit, see top right panel of Figure \ref{galfit}. The inner component has a  S\'ersic index n = 1.4 and an effective radius of 1.58{\arcsec}. The outer component has an effective radius of 4.9{\arcsec}. The flux ratio between the inner and outer components, i.e., bulge to disk ratio (B/T), is 2.36.

 Within the spatial resolution limit of the CFHT image, the presence of the central nucleus is not obvious, however, the addition of a central gaussian component in the {\sc galfit}  modeling slightly improves the fit in the inner region. The gaussian component light contribution is less than 1/50 of the bulge.  A careful inspection of the unsharp-masked image and the model-subtracted residual images seems indicate the presence of an inner bar/disk-like structure. The ellipticity profile along the major axis shows a mild change and the maximum ellipticity coincides with the maximum change in position angle.

The global photometric parameters (e.g., half-light radius and total luminosity) are  derived using a non-parametric approach. We follow a similar procedure as described in \cite{Janz08} where the fluxes are summed up within a Petrosian aperture. The K-band luminosity is derived from the 2MASS   
\footnote{http://irsa.ipac.caltech.edu} archival images. The bulge luminosity is derived using the inner and outer component flux ratio. Since the detailed decomposition has been done in CFHT $i-$band image, here we assume that the stellar population in cE\_AGN is homogenous. The results of photometry and structure modeling are listed in Table \ref{htb}.

 \begin{figure*}
   \includegraphics[width=18cm]{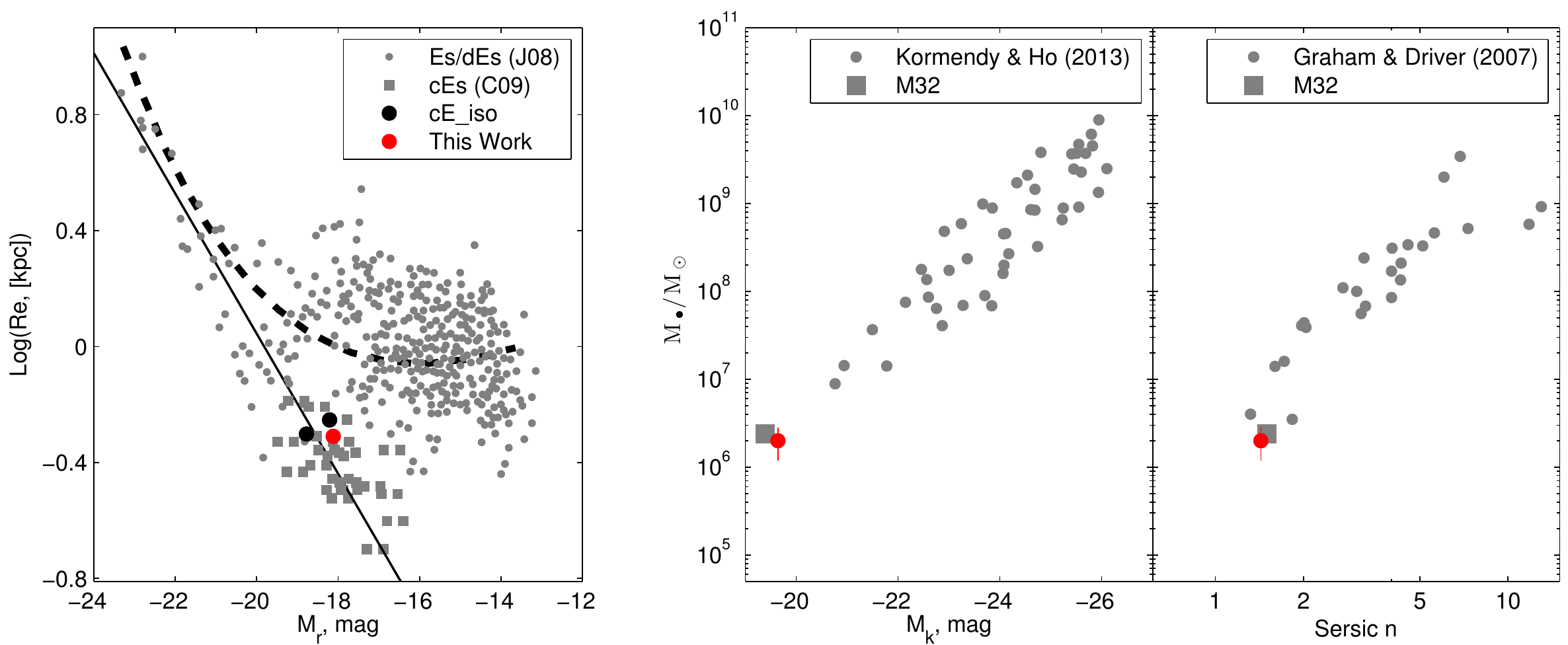}
  \caption{Left: Relation between Log(Re) and M$_{r}$. The plot is reproduced from P14, see their Figure 4. The additional point, cE\_AGN,  is shown in red. Here, we used the global photometric parameters, i.e. half light radius and total brightness. Right: Relations between BH mass and host galaxy bulge properties where \cite{Kormendy13} and  \cite{Graham07} samples are used to show the correlation between BH mass and bulge K-band magnitude and BH mass and bulge S\'ersic index, respectively. }
  \label{scl}
   \end{figure*}

 \subsection{Spectroscopy}

The SDSS spectroscopic observation of cE\_AGN was done as part of the BOSS\footnote{https://www.sdss3.org/surveys/boss.php} survey. The redshift measured by the SDSS pipeline is z = 0.0140. We exploit the optical spectrum of cE\_AGN for twofold. First, we aim to derive the properties of the underlying old stellar population using the information that is available in absorption features. Second, we measure the mass of the central BH by analyzing the stellar continuum subtracted emission lines. For this, we first fitted the observed galaxy spectrum in a wavelength range from 4000 to 7000 \AA{} with the Simple Stellar Population (SSP) models from \cite{Vazdekis10}. In doing so, prominent emission line regions were masked. For this purpose, we used a publicly available full spectrum fitting tool ULySS\footnote{http://ulyss.univ-lyon1.fr} \citep{Koleva09}.  An additive continuum of third order was added to account for  a non-thermal continuum while matching the galaxy and SSP spectra. The quality of fit is shown in the lower panel Figure \ref{sspfits}, where the  observed spectrum matches within 5\%  of the flux; excluding the masked emission-line regions. The best fit SSP parameters are Age = 9.95 Gyr and log(Z/Z$_{\sun}$) = $-$0.12 dex. 

Finally, we reanalyzed the residual spectrum that mainly contains emission lines (Figure \ref{sspfits} lower panel). The emission lines were de-blended into a narrow and a broad component by fitting multi-Gaussian models. A single Gaussian was used to model the narrow component of H$\alpha$, assuming the narrow component has the same profile and redshift as the [NII] doublet as shown in the inset of Figure \ref{sspfits}. However, for the [O III] doublet ($\lambda$ 4959 and $\lambda$ 5007) and the narrow component of H$\beta$, we adopt double Gaussians to account for the blue wing \citep{Greene05}. For the broad components  we use multiple Gaussians, which yields FWHMs of 1550, 2400 km/s for the broad H$\beta$ and H$\alpha$  components, respectively. Finally we measured H$\alpha$ and H$\beta$ narrow and broad emission fluxes from the best-fit narrow and broad component lines, respectively.

 Having carefully measured the H$\alpha$ and H$\beta$ flux from stellar continuum subtracted spectrum, the central black hole mass is measured by using the calibration provided by Eq. 4 in \cite{Ho15} where we replace H$\beta$ FWHM with H$\alpha$ FWHM and L5100 is converted from the L(H$\alpha$), using the conversion factor in \cite{Greene05}. The different normalization factors $f$ are assumed for classical bulge and pseudo bulge galaxies. With a S\'ersic index $n = 1.4$, cE AGN may be neither a de Vaucouleurs bulge ($n = 4$) nor an exponential disk ($n = 1$). Therefore, we used the scaling factor that was derived from combining both types of galaxies. The derived BH mass is $2.1 \times 10^6 M_{\sun}$.  The uncertainty for the BH mass measurement is approximately 0.4 dex which is mainly due to the uncertainty in the scaling factor.

\section{Scaling Relations}

The distribution of sizes and luminosities for early-type galaxies is shown in the left panel of Figure \ref{scl} left. This plot is reproduced from \cite{Paudel14} where the data points were compiled from \citet[][J08]{Janz08} for  dEs, Es and S0s, and \citet[][C09]{Chilingarian09} for cEs. The solid line is a power-law fit to the size and luminosity within the brightness range $M_{r} < -19.2$ mag. The dashed line is derived from the observed correlation between the central surface brightness and (global) S\'ersic index with luminosity, and this, not surprisingly, provides a common scaling relation for Es and dEs \citep{Graham03}, albeit showing significant systematic offsets at intermediate luminosities \citep{Janz08}. The two solid black circles are the isolated cEs previously identified in \cite{Paudel14}.

The red symbol is newly added from this work. We can clearly see that cE\_AGN overlaps with the previously identified cEs (black dots and solid gray squares). cE\_AGN falls significantly below the curved size-luminosity relation of Es/dEs, and rather continues the sequence of giant ellipticals shown by a solid black line. In other words, with the high mean surface brightness, $<\mu_{r}> $ = 19.36 mag/arcsec$^{2}$ for its total brightness of M$_{r}$ = -18.08 mag, it also follows the so-called Kormendy relation (magnitude vs surface brightness) of giant ellipticals \citep{Kormendy77}.

In the right panel of Figure \ref{scl}, we show the two scaling relations between the BH masses and the host galaxies properties. In the left, a well-known correlation between BH mass and K-band bulge luminosity of the host galaxy is shown \citep{Kormendy13}. In addition to this, a relation between BH mass and S\'ersic index, as proposed by \cite{Graham07}, is shown in the right panel. In both plots, cE\_AGN seems to roughly follow the observed trends, and interestingly, it almost overlaps with M32 which is shown by a large filled-gray square.

\section{Discussion and Conclusion}
We present a new compact early-type galaxy, cE\_AGN with optical signature of an accreting BH. Among the cE or dE class galaxies, cE\_AGN is the first which possesses the broad-line emission of width larger than 2000 km/s. Below, we discuss its distinction and the possible origin.

\subsection{Black holes in low mass early-type galaxies}
 AGNs are not common in low-mass early-type galaxies (dE or cE classes). According to a common wisdom these galaxies are devoid of gas and have less chance of acquiring gas from external sources as massive galaxies do, thus there is no fuel left to trigger the nuclear activity. Nevertheless, the number of detections of BHs in low-mass galaxies has been increased in recent literature with the advent of large scale surveys, high precision instruments and the sophisticated modeling of the host galaxy and BH kinematics \citep[e.g.,][]{Brok15,Greene04,Seth14}.

\cite{Barth04} identified a dwarf Seyfert 1 AGN candidate, POX 52, at redshift z = 0.021.  They classified the host galaxy morphology to be a dwarf elliptical and reported a BH mass of 1.5 $\times$ 10$^{5}$ M$_{\sun}$. In the last few years, number of searches have been devised to list  dwarf galaxies with central massive BHs \citep{Greene04,Greene07,Moran14,Reines13}. In particular, the study of \cite{Reines13} presents the largest number of nearby dwarf galaxies which exhibit optical spectroscopic signatures of accreting BHs. Visually classifying these BH-host dwarf galaxies located within z $<$ 0.02\footnote{We have chosen this redshift cut because beyond this the visual classification of dwarf galaxies from the SDSS color possibly is not reliable.}, we find that 13 out of 33 can be classified as dE or dS0. Only one, J122342.82+581446.4, has a broad H$\alpha$ emission similar to that of cE\_AGN. \cite{Reines13} estimate its  BH mass  $\sim$10$^{6}$ M$_{\sun}$.

\subsection{Origin of cEs}
 The majority of cEs are found in the vicinity of giant galaxies, e.g., M32, and many of them possess morphological features which might have originated from tidal stripping activity \citep{Huxor11}. With the discovery of an isolated cE by \cite{Huxor13}, an alternative scenario, other than stripping, has gained much attention in the recent literature \citep{Dierickx14,Paudel14}. cE\_AGN is also located away from the nearest massive galaxy. Recently, \cite{Chilingarian15} proposed a fly-by scenario where those isolated compact ellipticals may be tidally-stripped systems that ran away from their hosts.

The discovery of a cE with AGN signature might help us to explore another aspects of compact early-type galaxies. Bulge-dominated early-type galaxies follow a scaling relation between bulge mass and BH mass. Measuring the BH mass allows us to locate the position of cEs  in this relation. We find nothing uncommon, as  cE\_AGN is placed near M32, following the overall relation between BH mass and bulge mass (see Figure \ref{scl} right). M32 is a Local Group cE located at the vicinity of M31 which has been an iconoclast example of cE type galaxy \citep{Faber73}. Based on X-ray and radio observations, presence of the nuclear activity has also been reported in M32 \citep{Ho03}. It has been long argued that M32 is a stripped bulge of a disk galaxy \citep{Choi02,Faber73,Graham02} -- but also see \cite{Dierickx14}. As a hypothesis proposed by \cite{Chilingarian15}, the possibility that cE\_AGN might also be a tidally stripped bulge that ran away from its hosts, perhaps from NGC 2672 group, can not be ruled out.  NGC 2672 is a relatively massive galaxy group with a viral mass of $\sim$10$^{13}$M$\sun$ \citep{Ramella02}. It is expected that the member galaxies could be spread out as far as a couple of Mpc in projected distance \citep{Sales07}. However, given the large physical separation between cE\_AGN and NGC 2672, the stripping event might have happened either very early in time or at fast speed with a very eccentric orbit. The small relative velocity to NGC 2672 might be explained by an orbit in the plane of the sky and cE\_AGN being close to apogalacticon.

Other hand, It is remarkable to see that while following the extension of mass--size relation of massive early-type galaxies, cE\_AGN also seems to follow the observed empirical relation between the bulge S\'ersic index and the BH mass suggested by \cite{Graham07}. Interestingly, a detailed decomposition of the light profile by \cite{Graham02} reveals that M32's light profile resembles an almost perfect bulge - exponential disk system where the bulge component has a S\'ersic index of $n = 1.5$, which is in agreement with the relation between  the S\'ersic
index and the BH mass.  Considering the location in the scaling relations,  we suggest cE\_AGN is a low luminosity extension of Es. It is, though, not sufficient to rule out the stripping scenario and to support the alternative scenario which assumes a similar evolutionary path as Es, formed possibly via mergers \citep[e.g.,][]{Naab07}. Furthermore, we expect that a detailed study of the underlying stellar population and central kinematics from high quality 2-dimensional spectroscopy may allow us to test the above conclusions which we plan to do for the coming observing semester.

\acknowledgements
We thank Alister Graham for fruitful discussions and comments on the draft version of this paper. We also thank the referee, Igor Chilingarian, for helpful comments that improved the paper. This study is based on the archival images and spectra from the Sloan Digital Sky Survey (the full acknowledgment can be found at http://www.sdss.org/collaboration/credits.html).  We also made use of archival data from Canada-France-Hawaii Telescope (CFHT) which is operated by the National Research Council (NRC) of Canada, the Institute National des Sciences de l'Univers of the Centre National de la Recherche Scientifique of France and the University of Hawaii.

\end{document}